\documentclass{llncs}

\usepackage{graphicx}
\usepackage[pdftex,colorlinks=true]{hyperref}
\usepackage{url}
\usepackage{multirow}
\usepackage{pdfsync}
\usepackage{cite}
\usepackage{listings}
\usepackage{subcaption}
\usepackage{booktabs}
\usepackage{amsfonts}
\usepackage[usenames,dvipsnames]{xcolor}

\lstdefinelanguage{JavaScript}{
  keywords={typeof, new, true, false, catch, function, return, null, catch, switch, var, if, in, while, do, else, case, break},
  keywordstyle=\color{blue}\bfseries,
  ndkeywords={class, export, boolean, throw, implements, import, this},
  ndkeywordstyle=\color{darkgray}\bfseries,
  identifierstyle=\color{black},
  sensitive=false,
  comment=[l]{//},
  morecomment=[s]{/*}{*/},
  commentstyle=\color{purple}\ttfamily,
  stringstyle=\color{red}\ttfamily,
  morestring=[b]',
  morestring=[b]"
}

\usepackage[textsize=footnotesize]{todonotes}

\newcommand{\timesr}{\textcolor{red}{{\times}}}
\newcommand{\checkmarkr}{\textcolor{OliveGreen}{{\checkmark}}}

\begin{document}
\frontmatter 

\title{The Entity Registry System: Implementing 5-Star Linked Data Without the Web}

\author{ 
Marat~Charlaganov\inst{1} \and Philippe~Cudr\'e-Mauroux\inst{2} \and Cristian~Dinu\inst{1} \and 
Christophe~Gu\'eret\inst{1} \and Martin Grund\inst{2} \and Teodor~Macicas\inst{2}\thanks{Authors are listed in alphabetical order.}
}
 
\institute{
DANS, Royal Dutch Academy of Sciences---The Netherlands \\
\email{\{firstname.lastname\}@dans.knaw.nl}
\smallskip
\and
eXascale Infolab, University of Fribourg---Switzerland\\
\email{\{firstname.lastname\}@unifr.ch}
}
\authorrunning{Charlaganov et al.} 

\maketitle

\begin{abstract} 
Linked Data applications often assume that connectivity 
to data repositories and entity resolution services are always available. This 
may not be a valid assumption in many cases. Indeed, there are about 4.5 
billion people in the world who have no or limited Web access. Many data-driven 
applications may have a critical impact on the life of those people, but are 
inaccessible to those populations due to the architecture of today's data 
registries. In this paper, we propose and evaluate a new open-source system that can be used as a 
general-purpose entity registry suitable for deployment in poorly-connected or 
ad-hoc environments. 
\end{abstract}
  
\section{Introduction}
\label{sec:Introduction}

There is an estimated number of 2 billion individuals that have access to the
Internet and can thus use centralized cloud hosted solutions for sharing data.
Many of these centralized solutions are well-known (Facebook, Wikipedia, WikiData, etc.) and
make it possible to share semi-structured data about online entities. Unfortunately, the populations who do not
have seamless data connectivity cannot rely on such data sharing
solutions even if they have computers that are interconnected through local mesh networks.

The OLPC (One-Laptop-Per-Child) initiative\footnote{\url{http://one.laptop.org/}} is bringing Information and
Communication Technology (ICT) to young learners in the poorest areas of the world
so that necessitous children can benefit from using ICT tools to develop new skills
too and from working collaboratively using
multi-media applications. So far, two million children
world-wide have been introduced to ICT by the OLPC foundation. Studies have shown
that such programs lead to an increase of the children's problem solving capabilities and
general computer skills.

One important technical problem remains, however: all the data created by the children on their learning
devices (\textit{e.g.}, OLPC XO-1) stays on the device. The devices are most often used in a
closed network, disconnected from the Internet, and all sharing mechanisms are
synchronous. In addition to the XOs, there is an increase in the amount of
low-resource computing devices (PlugPCs, tablets, etc.) that are made available to
increase the ICT reach to those that currently cannot benefit from it.

Due to the broader availability of devices, new problems arise. Even if everyone gets an
Internet-enabled device, not everyone will actually get seamless access to the Internet 
in developing countries. Most notably, there are even some cases where countries are deliberately
cut from the Internet for political or cultural reasons.

Accelerating the adoption of Linked Open Data (LOD) and data-intensive applications
in developing regions  with limited 
Internet connectivity hence requires adaptation to the specific challenges posed by the
living conditions of those 5 billion world citizens not having constant Internet
access. Among several challenges we previously highlighted~\cite{Gueret2011}, the
Entity Registry System (ERS) project tackles the design of an entity registry that can be globally edited
using a swarm of small devices interconnected in an intermittent way.
The goal of this registry is to replace the Web as a platform to
publish linked data whenever the latter is not available. The ERS allows linked data to be put \emph{into use} 
in the many regions where Internet connectivity is not guaranteed.

Providing and consuming linked data without the Web yields a number of issues.
In this work, we focus our efforts around the following questions:
\begin{itemize}
  \item How can non-colliding unique identifiers for resources be minted in a 
   un-coordinated way?
   \item How can data accessibility be ensured when the original data host is 
   offline?
 \item What is the best internal representation for entity descriptions made of
   the contributions from several nodes?
   \item How can the registry storing all entity data be loosely-coupled to the Web?
   \item How can one optimize transactional synchronization for loosely-connected
   devices?
  
\end{itemize}

This project investigates these questions in the context of the use of LOD principles
on XOs and PlugPCs having intermittent Internet connectivity. We have implemented and
open-sourced the ERS system that enabled data sharing in mixed environments, and are
now testing and deploying the system in the context of several OLPC initiatives. 

The rest of this paper is organized as follows. In Section~\ref{sec:Sharing}
we introduce the problem of data sharing and the different technical solutions
to it. This section is followed by Section~\ref{sec:ERS} where the Entity
Registry System (ERS) is introduced along with a reference implementation (Section~\ref{sec:Implementation}). 
We analyze the performances of the reference implementation in Section~\ref{sec:Performance}
and conclude in Section~\ref{sec:Conclusion}.

\section{Sharing structured data}
\label{sec:Sharing}

We consider the following task: a user wants to associate a number of properties
to an entity identified with a unique identifier. This description has to be
open and accessible to other users (read only or read+write depending on the situation).

The most common approach to process such a task nowadays consists in setting up a centralized server, backed up
with a cluster for scalability. Every time the application has access to that
server it can modify the content of the database, creating and deleting entities
as well as updating their description. While this is by far the most popular
approach, its main drawback is the loss of functionality as soon as the
connection to the central server becomes unavailable.

A second approach consists in adding a temporary offline storage and special
synchronization routines on the client (\textit{c.f.}, for instance, Mendeley 
or Evernote desktop applications). In that way, a temporary loss of connectivity 
can be compensated by the use of cached data and the synchronization of edits 
as soon as the connection returns. This approach is particularly suitable when 
the loss of connection is an exceptional event and when only when a handful of 
individuals have access to the data. It is facilitated by recent 
developments of W3C recommendations (\textit{e.g.}, Web Storage). This approach is impractical
as soon as
disruptions are more frequent, or are the default mode, or when more people
need to have access to the data.

Some systems assuming a perpetual un-connected state, \textit{e.g.}, many former-generation GPS navigation
systems that solely rely on their local database. This leads to a third approach
that consists in a locally accessible data source whose availability is
guaranteed locally and can be made available sporadically through ad-hoc network access. The main drawback of this approach is that
it is then much harder to expose the local data to other systems (\textit{cf.} PageKite\footnote{\url{http://pagekite.net/}} ).

\begin{figure}[h!tbp]
\centering
\includegraphics[width=0.7\linewidth]{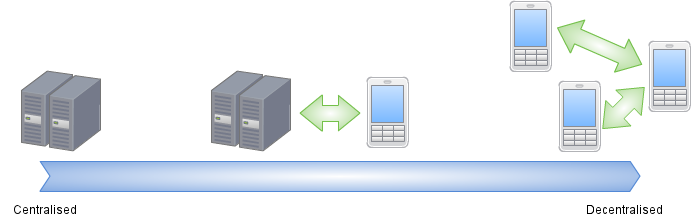}
\caption{The evolution from fully centralized solution to fully decentralized 
alternatives}
\label{fig:decentralisation}
\end{figure}

Figure~\ref{fig:decentralisation} gives a simple pictorial comparison of those
three approaches. Up to now, the vast majority of linked data solutions 
were built following a centralized
paradigm: the data is put in one location and accessed by a client
application hosted at another location. As mentioned above, this system
architecture fails as soon as the connection becomes unavailable.

With ERS, we aim to provide a local system that can also operate in connected
settings. In our approach, we want to seamlessly transition from decentralized
to centralized settings depending on the current connectivity. We
model a group of XOs/PlugPCs/Tablets/etc. as a swarm of devices that operate
in a finite context. Even though Internet connectivity is
limited, we assume devices can directly communicate (\textit{e.g.}, through a 
mesh network or an infrastructure network in a class room). 

In the following, we first give an overview of the system in Section~\ref{sec:ERS}, and then 
describe our reference implementation in Section~\ref{sec:Implementation}.

\section{The Entity Registry System (ERS)}
\label{sec:ERS}

The Entity Registry System (ERS) for storing semi-structured descriptions of 
entities is designed around lightweight components that collaboratively support 
data sharing and data-intensive applications in intermittently connected settings. It is
compatible with the RDF data model and makes uses of both Internet and local networks
to share data, but does not base its content publication strategy on the Web. No
single component is required to hold a complete copy of the registry. The
global content consists of the union of what every component decides to share.


\subsection{Components}
ERS is articulated around three types of components: Contributors,
Bridges, and the Global Server. The components can be deployed on 
any kind of hardware ranging from low-cost computing devices such 
as the RaspberryPi\footnote{\url{http://www.raspberrypi.org/}} to 
fully-fledged data centers.

\subsubsection{"Contributor" component}
Contributors read and edit the contents of the registry. They may 
create and delete entities, look for entities, and contribute to the
description of the entities. Every contribution made by a contributor
is identified by its name within the system so that the collectively-created
description of an entity can be traced back to individual contributions.
Contributors are free to make any statement about any entity in the system.
They use a local data-store in which they persist the description of
the entities.

Contributors can also cache the contribution of others. In particular, a data replication 
mechanism ensures that the descriptions of entities relevant to a contributor 
are made available on his own store. For example, a contributor storing some 
properties for an entity \verb|X| will trigger a synchronization process for 
getting everything everyone said about \verb|X| for local use.

The identifier for the entities can be freely picked by developers using ERS. 
The only constraint is that this identifier must be a URN and contain
a path, as follows: \texttt{urn:ers:<path>:<identifier>}. The path can be a UDC
class\footnote{\url{http://www.udcc.org/}}, the name of the software that created the entity, 
a FQDN from the LOD. Part of the goals of the
project is to investigate how some URNs end up being preferred over others as
part of a collaborative reinforcement process similar to the one driving data sets
re-use on the LOD~\cite{DBLP:conf/www/MaaliCP11}. We choose URNs because the identifiers
are not meant to be resolvable outside ERS, an optional "global server" can establish
the connection between ERS and the Web of Data.


\subsubsection{"Bridge" component}
Bridges do not directly contribute to the content of the registry. They are used to 
connect isolated closed networks and improve the availability of the individual 
descriptions shared by the contributors. Bridges can theoretically store content 
coming from any contributor, but will typically store the data only for a limited 
amount of time (\textit{e.g.}, using \emph{soft states}) due to their limited 
capacity. To summarize, bridges have two core functionalities: distribution of 
local data in semi-connected networks and synchronization to other closed networks.


The synchronization process with the bridges is simple and consists in i) sending 
to the bridge every new description from the contributors, and ii) getting back 
descriptions that are useful to the devices (that is, retrieving data for  
entities that the contributors have already considered or are requesting).

The major challenge for the bridges is to maintain the integrity of the data
they store. Therefore, the goal is to provide the right transaction mechanisms that
allow to maintain a consistent state of the data and the best possible
performance even when multiple contributors are editing and looking-up the entities. 
Another goal is to achieve a separation of concern by leveraging the knowledge about the different
contexts defined by the distinct contributor networks.

\subsubsection{"Global server" component}
ERS deployments can feature any number of bridges and contributors. In addition, 
some use-cases may require the presence of a global server that contains a copy 
of all the data provided individually by the contributors. The global server 
provides a single, read-only, entry point to the registry. It exposes the contents 
of the ERS to other systems, for instance to the Web of Data. This global server 
is connected to the bridges and keeps copies of everything that transits through 
them. Like the bridges, the presence of this global server is optional.

When used, the global server harvests data from the bridges. The server 
aggregates the individual contributions every contributor made to the
entities and expose this information as an integrated, 
read-only, dataset. The main challenge for the global server is scalability 
(\textit{c.f.} Section~\ref{sec:Performance}).


\subsection{Usage}
\begin{figure}[h!tbp]
\centering
\includegraphics[width=0.6\linewidth]{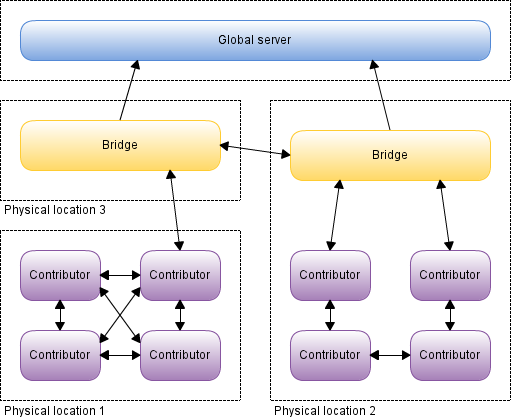}
\caption{An example ERS deployment across three different locations}
\label{fig:ERS}
\end{figure}

Figure~\ref{fig:ERS} shows and example deployment featuring three
different physical locations, 8 contributors, two bridges, and a global server. 
The contributors are devices creating, consuming and storing structured data 
about entities. One bridge is used to ensure information flow and data 
distribution between the nodes of ``physical location 2'' (exhibiting poor connectivity) and those of
``physical location 1''. The second bridge is located in a 
separated network where both physical locations can connect to. In this 
scenario, the second bridge is used for synchronization and data flow between 
the two separated networks. The global server is used to expose the entities within
ERS as de-referencable HTTP URIs.

Concretely, ERS can for instance be used for asynchronous sending of messages 
between XO laptops in schools or collaboratively editing a multi-lingual
corpus.

\begin{description}
\item[Mailing application]

The mailing application lets every XO having ERS installed on it send and 
receive mails with any other XOs. To send a mail, the software creates a new 
entity and associates to it the message, the name of the creator and the name of 
the target device. This description gets then automatically replicated to the 
target device, eventually transiting through a bridge. On the receiver side, the 
messaging application only has to display in 
the receive box every entity description that has been targeted for this device 
but created by another device. Links between messages can be established by 
referring to the unique identifiers of the entities.

\item[Multi-lingual tagging]

With the increase of connectivity and the pressure towards the uniformization of 
communications, there is a need for crowdsourcing language preservation and 
cultural heritage~\cite{icend2011} 
In this 
scenario, ERS is used to power up a social gaming where users can tag items 
with the name of that item in their own language and also connect items to each 
others to group them. A user can play that game at School, take the game back 
home and continue playing it with his parents, and then return to school to 
share the new results with his peers. ERS also makes it easy to gather the 
globally crowdsourced content by monitoring the content of bridges or using
a global server.
\end{description}

\section{Reference implementation}
\label{sec:Implementation}

In this section, we give a detailed overview of the reference implementation
of ERS. All of the concepts we mention in this paper are implemented and
available as open source code from the ``ers-devs'' group on
GitHub\footnote{\url{https://github.com/ers-devs}}. The following description is 
organized around three subsections, each of them describing one of our components
in more detail.

\subsection{Contributors}
We choose to use CouchDB\footnote{\url{http://couchdb.apache.org/}} to persist 
all the data locally and perform the different synchronizations. The main advantage of 
CouchDB is that it provides built-in mechanisms for flexible replication of the 
data using the ad-hoc mesh network capabilities of the XO laptops. The CouchDB 
replication system can be configured to match our needs (create triggers and 
filters based on the descriptions).

Internally, CouchDB defines \textit{documents} that are used to store and replicate 
payloads of JSON data. There are different encoding choices that can be made to 
use this system to store RDF data. Taking our inspiration from previous work 
(
LD-In-Couch\footnote{\url{https://github.com/mhausenblas/ld-in-couch}}), we 
experimented with different storage representation.

\subsubsection{Storing of predicate/object pairs}
The two options here are reified statement or array of values. We want to allow 
several values to be associated with the same property. As properties are 
unique in CouchDB, this is only possible by either storing all the values as an 
array or storing an array of properties coupled with an array of values.
Let us consider the following N-Quads:
\begin{verbatim}
ers:message1 ers-prop:body "hello world" ers:xo1 .
ers:message1 ers-prop:to ers:xo2 ers:xo1 .
ers:message1 ers-prop:to ers:xo3 ers:xo1 .
\end{verbatim}

These can either be expressed using the predicate as keywords 
(Listing~\ref{lst:json1}) or two synchronized arrays 
(Listing~\ref{lst:json2}).

\begin{tiny}
\begin{lstlisting}[language=JavaScript,frame=single,label=lst:json1,caption={
JSON Model 1 - use predicate as key}]
{
  "_id": "message1 xo1",
  "body": ["hello world"],
  "to": ["xo2", "xo3"]
}
\end{lstlisting}
\end{tiny}

\begin{tiny}
\begin{lstlisting}[language=JavaScript,frame=single,label=lst:json2,caption={
JSON Model 2 - use two synchronised arrays of p/o values}]
{
  "_id": "message1 xo1",
  "p": ["body", "to", "to"]
  "v": ["hello world", "xo2", "xo3"],
}
\end{lstlisting}
\end{tiny}

An estimate of disk space required for storing the standard $SP^2$bench 
dataset using these serialization models is given in Table~\ref{storage}.
For this experiment we switched off the database file compression in CouchDB. 
Disk usage is very similar in all three cases and thus not
a decisive factor.

\begin{table}[h!tbp]
\centering
\begin{tabular}{c|c}
\textbf{Serialisation} & \textbf{Size (kB/1K triples)} \\
\hline
N-Triples & 157 \\
Model 1 & 143 \\
Model 2 &  157 \\
\hline
Document per quad & 418 \\
\end{tabular}
\caption{Storage requirements of different serialization models.}
\label{storage}
\end{table}

\subsubsection{Usage of documents}
We also considered several options for storing entities using CouchDB:
\begin{enumerate}
 \item \textbf{Create one document per entity}. Every contributor will be 
updating the same document associated to a particular entity he/she wants to edit. 
This approach gives good read performance, as all the data needed
by a contributor interested in the entity is available in one go. However, synchronization
becomes extremely problematic because documents must be merged during replication
while accounting
for users with different access rights editing various parts of a document at different times.
CouchDB does not have out-of-the-box support for such features (it can only replicate documents without
transformation). Moreover, some entities can grow very large despite only a small portion of
them ever being queried or modified. This leads to inefficient I/O and disk space usage,
especially since CouchDB updates by appending new versions at the end of the documents.

\item \textbf{Create one document per contribution made to the description 
of an entity}. In RDF terms, this translates to one document per entity x graph
combination. The synchronization process only has to ensure documents are 
made available \textit{as is} to the nodes that need them, since in our system,
a graph can only have one author and one set of R/W permissions, and thus the
potential for conflict is essentially eliminated. I/O and disk space are better used, as only
portions of a entity that have changed are transmitted. Read performance, however, will be
worse because a client will now have to do several lookups to fetch and load all the documents 
describing a particular entity.

\item \textbf{More granular approaches}, such as one document per graph-entity-property
combination, or even one document per quad. These offer little benefit, as the I/O and storage overhead
as well as the read performance quickly become unacceptable.
\end{enumerate}

We performed several experiments to help us pick the best solution, and in the end chose \textbf{one document per graph-entity combination} as the best of both worlds in terms of
complexity and performance.

\subsection{Bridges}

We now describe the intermediate level in our global system architecture, the Bridges.

Implementation-wise, Bridges are actually very similar to Contributors. They use the same
internal data model and are also based on CouchDB. The main difference is in the 
contributor-bridge and bridge-bridge synchronization rules. The three situations are
presented here for comparison:

\subsubsection{Contributor-Contributor Synchronisation}

Apart from their locally generated data, contributors only accept annotations that are
specifically addressed to them (by marking the corresponding graph with a specific property). Contributors
make public all data that is not explicitly marked as private. Contributors also temporarily cache the results
to recent queries. Therefore, through either replication or querying its neighbors, a contributor has access to:

\begin{enumerate}
\item The public data authored by the neighboring contributors
\item The data specifically addressed to it by any contributor in the system
\item The recently accessed data (if the query is made public) of all the contributors in the neighborhood.
This effectively produces a distributed cache that greatly accelerates access to common data.
\end{enumerate}

\subsubsection{Contributor-Bridge Synchronisation}

Unlike a Contributor, a Bridge accepts \emph{all} non-private data from nodes connected to it. All data in a Bridge
is made available to connected contributors. The Bridge does not contribute any data itself and is generally headless (\textit{i.e.} has no GUI for data input).

Note that a Bridge also accepts cached query data. Due to its larger storage capacity (which is needed anyway because it aggregates data from multiple Contributors), a Bridge can afford to maintain a larger cache for a longer period of time, and can thus act as a sort of second-level cache. Note also that no data on a Bridge is stored permanently - data that has not been needed for a long period of time is garbage-collected to make room for new information.

\subsubsection{Bridge-Bridge Synchronisation}

Bridges also share all data between them, with no authorship filters, except for the cached query data. This is because such data is generally only useful at a local level (\textit{e.g.}, within a classroom). A cache acting on a more global level would be too large, and this role is mostly fulfilled by the global server anyways.

\subsection{Global server}

The last element in our reference implementation is the so-called \emph{Global
Server}. While other elements in
the hierarchy are centered around sporadic or ad-hoc connectivity, a
\emph{Global Server} is an optional, always-on component. A single
instance of a \emph{Global Server} might not always store the entirety of data, but can
 store references to other servers providing such information, thereby
acting like a typical registry for non-local content. Relaying data from other
instances does not require caching on the server, since we assume the other
global servers will be always available.

Due to the distributed nature of how data is processed, cached and forwarded in ERS,
providing a consistent storage layer is not trivial. To achieve the desired
consistency, we implemented a transaction layer on top of our data store. Using
traditional distributed transaction schemes from relational database management
systems is out of scope for us for two reasons: First, allowing multi-side
distributed transactions would severely slow down the overall performance, and
second, even if the \emph{Global Server} is considered to be always available,
this is not true for contributing bridges. Thus, the goal must be to  achieve
the best possible throughput on slow dial-up connections even if the connection
is dropped multiple times.


The foundation for the storage layer of the \emph{Global Server} is a cluster of
Apache Cassandra nodes with an additional abstraction layer for storing RDF
inside the cluster based on CumulusRDF\cite{ladwig2011cldmonks}. While Cassandra
natively offers no support for transactions or atomic operations on data stored
in the cluster, we added support for atomic operations in the RDF layer.

\begin{figure}[h!tb]
  \centering
  \includegraphics[width=0.4\textwidth]{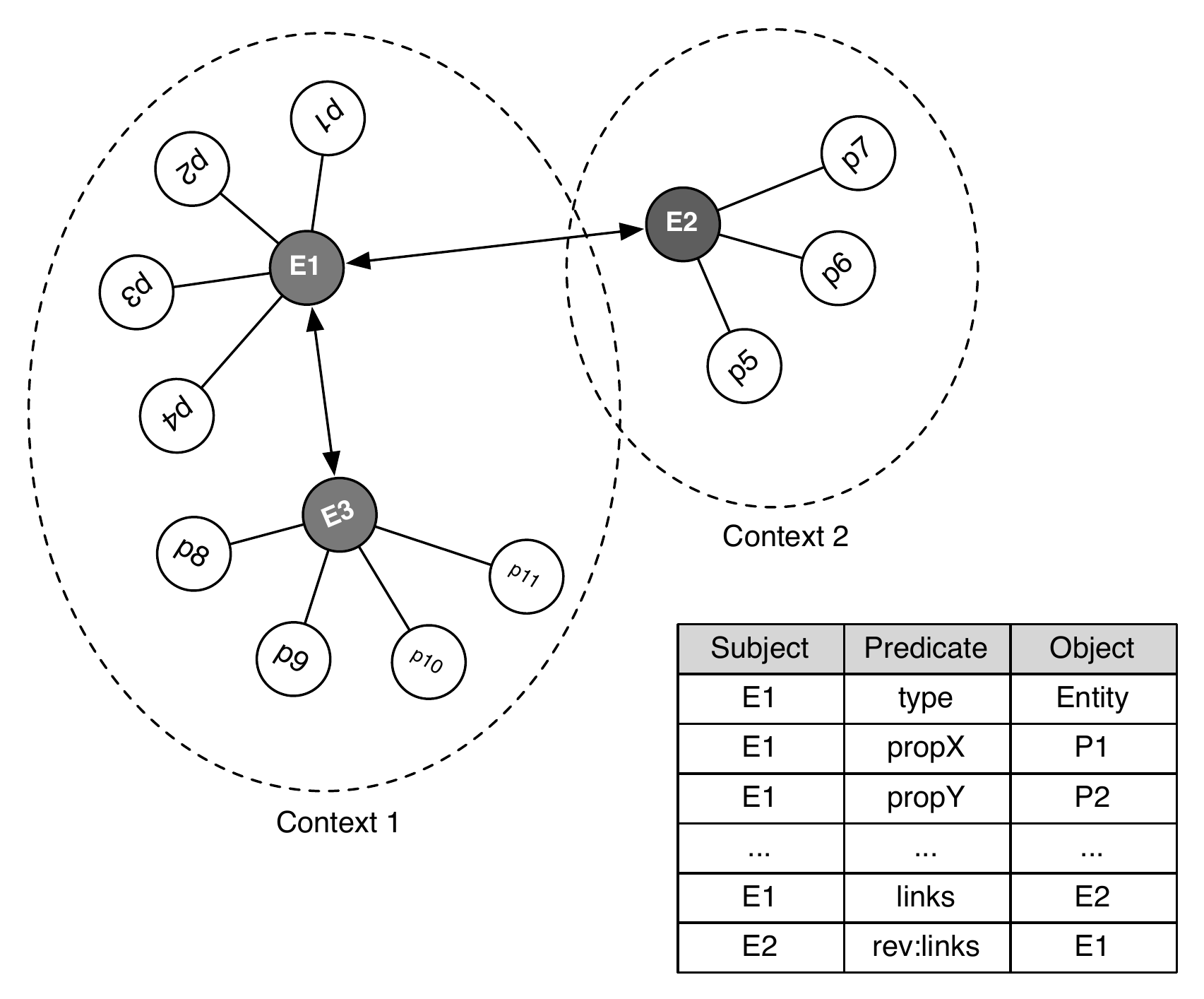}
  \caption{Entity Model Overview}
  \label{fig:data_model}
\end{figure}

The overall data model of the entity registry is shown in
Figure~\ref{fig:data_model}. There are two important semantic properties that
are visible in this figure. First, the properties of specific entities have no
connections between each other, this means that for a given triple $t=\{s,p,o\}$
with $s \in S, p \in P, o \in O$ there can be no two triples that share the same
predicate and object. For example, the case $T=[t_1=\{s_1,p,o\},t_2=\{s_2,p,o\}]$
is undefined because it violates this property, $t_1$ and $t_2$ sharing the same
predicate and object. This effectively describes a uniqueness constrain on all
triples on the $p$ and $o$ attributes. The second semantic property is that
 modifications on the graph of entities are typically separated by the
context in which they operate, making it less likely that collisions between
different contexts will occur. These contexts typically follow the same
structure as the different contributor/bridge landscapes.

Concerning links, only connections between entities are allowed. For simplicity,
a connection between an entity is seen as bidirectional. On the storage side we
model this as two entries in the graph, one describing the original connection
and a second entry describing the inverse relation. This allows us to navigate
the path in both ways and to keep the original semantics about how the
connection was created.

Inside our thin application layer sitting on top of Cassandra, we define the following
atomic operations:
   \emph{Insert entity (IE),
   Insert property of an entity (IP),
   Update property of an entity (UP),
   Delete property of an entity (DP),
   Delete entity (DE),
   Shallow entity copy (SC),
   Deep copy of an entity (DC),
   Insert bi-directional link between two entities (IL),
   Delete link between two entities (DL).}

Since transaction support must be implemented on a higher level we define a
multi-level locking scheme that allows hierarchical locking of the different
elements of an entity. In contrast to traditional relational databases our
locking approach has the possibility to lock an entity even if it does not exist
by referencing the unique ID of the entity in our lock table. This allows a
strict serialization of conflicting operations, even for insertions. The two
hierarchical locks are: $L_{E+P}$ and $L_{E}$. The former locks based on the
entity ID and the property, while the latter locks the complete entity. While
two $L_{E+P}$ locks can be compatible in case they differ in one of the two
parts, two $L_P$ locks are not compatible. Table~\ref{tab:tx_compat} shows the
compatibility for all kinds of operations with these different lock types. Since
all property locks are compatible we can achieve a high throughput for most of
the incoming operations. For the two lock types, we match the following
operations. Using the fine granular $L_{E+P}$ we can run the following
operations: IP, UP, DP, SC, IL, DL. Using the $L_E$ lock we can execute: IE, DE,
DC. For links and shallow copies the matching property is either \verb|sameAs|
or \verb|linksTo|.
We experimentally test the performance of such atomic operations in the following. 

\begin{table}[h!b]
  \centering
  \begin{tabular}{ccccccc} \toprule
                & $L_{E_a+P_c}$ & $L_{E_a+P_d}$ & $L_{E_b+P_c}$ & $L_{E_b+P_d}$ & $L_{E_a}$ & $L_{E_b}$\\
  \midrule
  $L_{E_a+P_c}$ & $\timesr$     & $\checkmarkr$ & $\checkmarkr$ & $\checkmarkr$ & $\timesr$ & $\checkmarkr$\\
  $L_{E_a+P_d}$ & $\checkmarkr$ & $\timesr$     & $\checkmarkr$ & $\checkmarkr$ & $\timesr$& $\checkmarkr$\\
  $L_{E_b+P_c}$ & $\checkmarkr$ & $\checkmarkr$ & $\timesr$ & $\checkmarkr$ & $\checkmarkr$ & $\timesr$\\
  $L_{E_b+P_d}$ & $\checkmarkr$ & $\checkmarkr$ & $\checkmarkr$ & $\timesr$ & $\checkmarkr$ & $\timesr$\\
  $L_{E_a}$     & $\timesr$     & $\timesr$     & $\checkmarkr$ & $\checkmarkr$ & $\timesr$ & $\checkmarkr$ \\
  $L_{E_b}$     & $\checkmarkr$ & $\checkmarkr$ & $\timesr$ & $\timesr$ & $\checkmarkr$ & $\timesr$ \\
  \bottomrule
  \end{tabular}
  \caption{Operation Compatibility}
  \label{tab:tx_compat}
\end{table}

In the spirit of web-scale NoSQL data stores we defer conflict resolution of
failed transactions to the application layer.


\section{Performance}
\label{sec:Performance}

This section reports on some testing we did to measure the performance of different
aspects of the system under various scenarios.

\subsection{Local Tests}

The first tests cover the local aspect of the system, which is to say the interaction between
Contributors and Bridges in particular.

\subsubsection{Experimental Setup}

For the experiments in this section we used 4 OLPC XO-1 configured as Contributor
nodes. The laptops run the local component of the ERS over CouchDB 1.2.1. 
The XO-1 hardware platform is characterized by 433 MHz x86 AMD Geode LX-700 CPU,
256MB total RAM, and 1024 MB of NAND flash memory. Software-wise, the laptops run a
custom desktop interface (Sugar) on top of Linux.

The laptops were configured in a mesh network that is typical of their most frequent
use case. Some experimental scenarios also include a Bridge node, in our case a Raspberry Pi,
which is a popular low-cost embedded computing platform fitted with an
ARM1176JZF-S 700 MHz CPU, 256MB of RAM and SD card storage (8GB). The Pi also runs Linux.

\subsubsection{Local Read/Write Performance Under Replication}

For our first experiment, we set up the XOs to use the ERS API to continuously write new random documents to
their internal store, while at the same time receiving the writes of other XOs in the mesh network
through replication, as well as performing queries in parallel. As an indication of the \emph{read performance}, we
measured the average latency for the queries within a 5-minute period. For measuring the \emph{write performance},
at the end of the 5 minutes, the total number of documents in the XO's stores was counted. The write performance
was then calculated as:

\[WritePerformance = \frac{TotalNumberOfDocuments}{NumberOfNodesInMesh \cdot ElapsedTime}\]

The number of nodes in the mesh was varied from 1 to 4, without a bridge node, undergoing all-to-all (mesh)
replication. One final scenario was added featuring 4 nodes with star replication via a Bridge.
The results are shown in Table~\ref{tab:storage}.

\begin{table}[h!tbp]
\centering
\begin{tabular}{c|c|c|c}
\textbf{Network Size} & \textbf{Repl. Type} & \textbf{Read Latency (sec)}& \textbf{Write Performance, docs/sec} \\
\hline
1 & all-to-all & 0.24 & 7.31 \\
2 & all-to-all & 0.36 & 4.65 \\
3 & all-to-all & 0.55 & 3.23 \\
4 & all-to-all  & 1.01 & 2.95 \\
4+bridge & star & 0.98 & 5.56 \\
\end{tabular}
\caption{Local Read/Write Performance Under Replication}
\label{tab:storage}
\end{table}

It can be seen that all-to-all replication has a significant effect on both read and
write performance. As the number of nodes
increases, the write performance converges to about 45\% of the 1-node case (which is equivalent to
no replication). However, the performance drop can be satisfactorily addressed through the use or star
replication via a Bridge node.

\subsection{Global server}
For the following experiments, we used 5 servers part of our cluster in
Switzerland (8-cores i7 Intel CPU, 8GB total RAM memory, Gigabit Ehternet, Linux
kernel 3.2.0, Java SE 1.6). All of them are running a Cassandra instance with
replication level 2. 

In the experiments, we varied different parameters impacting the overall
performance. We looked mainly at the overall throughput of our transaction
implementation but in addition varied the write consistency of the cluster
between ALL and ANY.

One of the machines runs the ERS Java program that was built on top of
cumulusRDF. It is the access point of our ERS system as well as the
central coordinator for transactional support. However, as the support is
implemented at the Java application layer and not at the Cassandra level, it is not yet
distributed. A future approach using ZooKeeper, Chubby or Cage is envisioned.
The clients were running on a different machine to better load balance and to
involve the network delay.

We use two different locking granularities as follows: for simple operations of
inserts, updates, and deletes we lock at the predicate (E+P) level, but for cloning operations we lock at
the entity level, since cloning operations would have involved more overhead to
iterate the entire transaction and lock every triples. Thus, a coarser lock is
used for this purpose.

In case a transaction has a lock contention conflict, it is aborted and
restarted for a maximum of 10 times before considering it an aborted transaction.

The total dataset size we used is about
10M triples with 1M unique entities and between 8-12 properties per entity. The
total size on disk is 2.7GB. 

\subsubsection{Throughput}

\begin{figure}[h!tb]
  \begin{center}
    \begin{subfigure}[b]{0.49\textwidth}
      \centering
      \includegraphics[width=\textwidth]{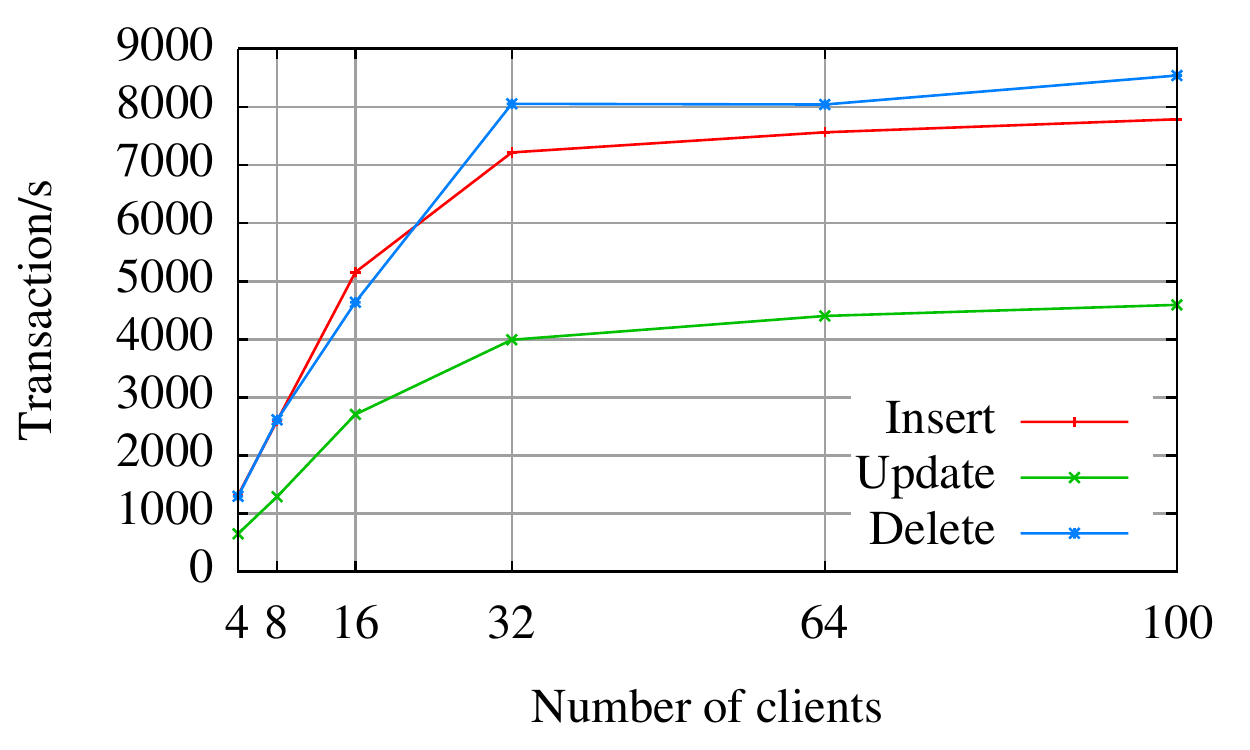}
      \caption{Basic Operations}
      \label{fig:basic_ops}
    \end{subfigure}
    \begin{subfigure}[b]{0.49\textwidth}
      \centering
      \includegraphics[width=\textwidth]{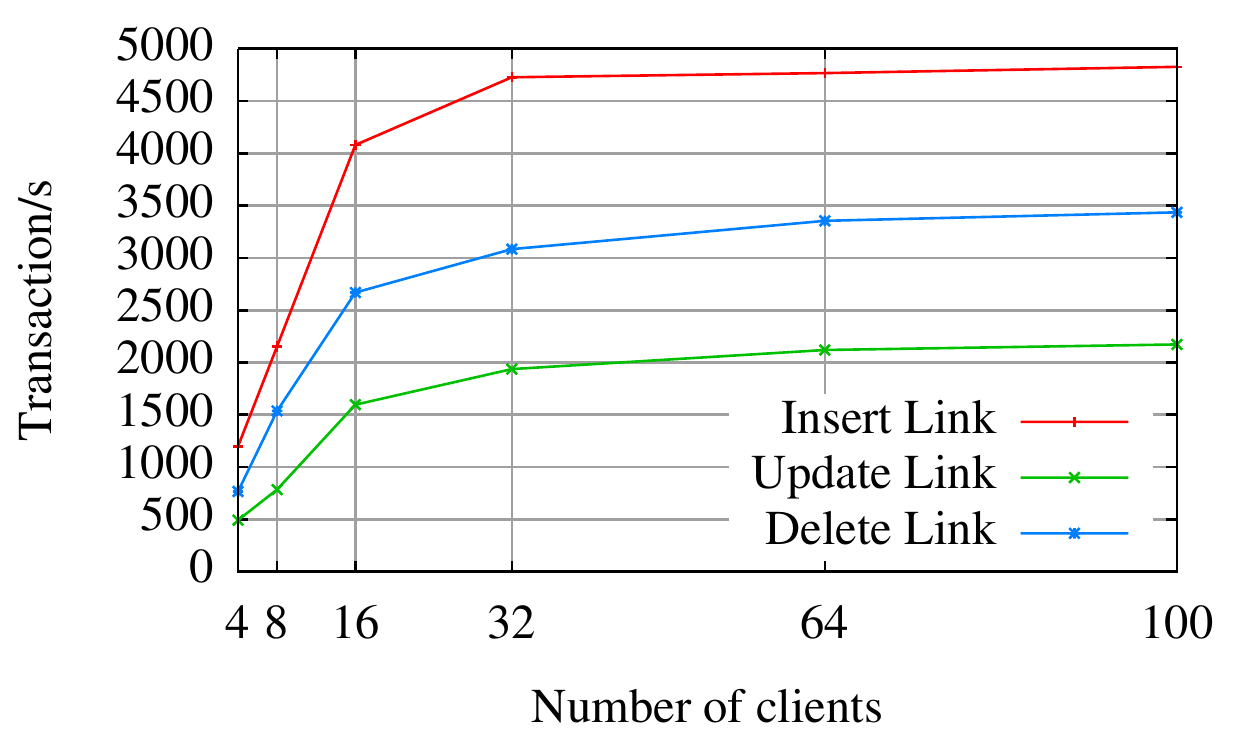}
      \caption{Linking Entities}
      \label{fig:linkin_ops}
    \end{subfigure}
  \end{center}
  \caption{Throughput for different operations in ERS. All transactions per
  clients are executed sequentially, but all clients run in parallel.}
  \label{fig:figure_tx_part1}
\end{figure}

At first, we analyze the overall throughput of the cluster depending on the
different operations we perform. We differentiate between basic operations that
 modify properties of existing entities and linking operations that
insert a bidirectional link between two existing entities. We use a pool with
a varying number of clients (2-64). For the basic operations shown in
Figure~\ref{fig:figure_tx_part1}, we can see that the throughput is maximized at
around 64 parallel clients. Delete operations have the highest throughput as
they only require simple marking of a record in the cluster with a tombstone by
Cassandra. For linking the situation is slightly different as it requires to
bundle two operations---two inserts---in one atomic operation, adding additional
overhead.

\begin{figure}[h!tb]
  \begin{center}
    \begin{subfigure}[b]{0.49\textwidth}
    \centering
    \includegraphics[width=\textwidth]{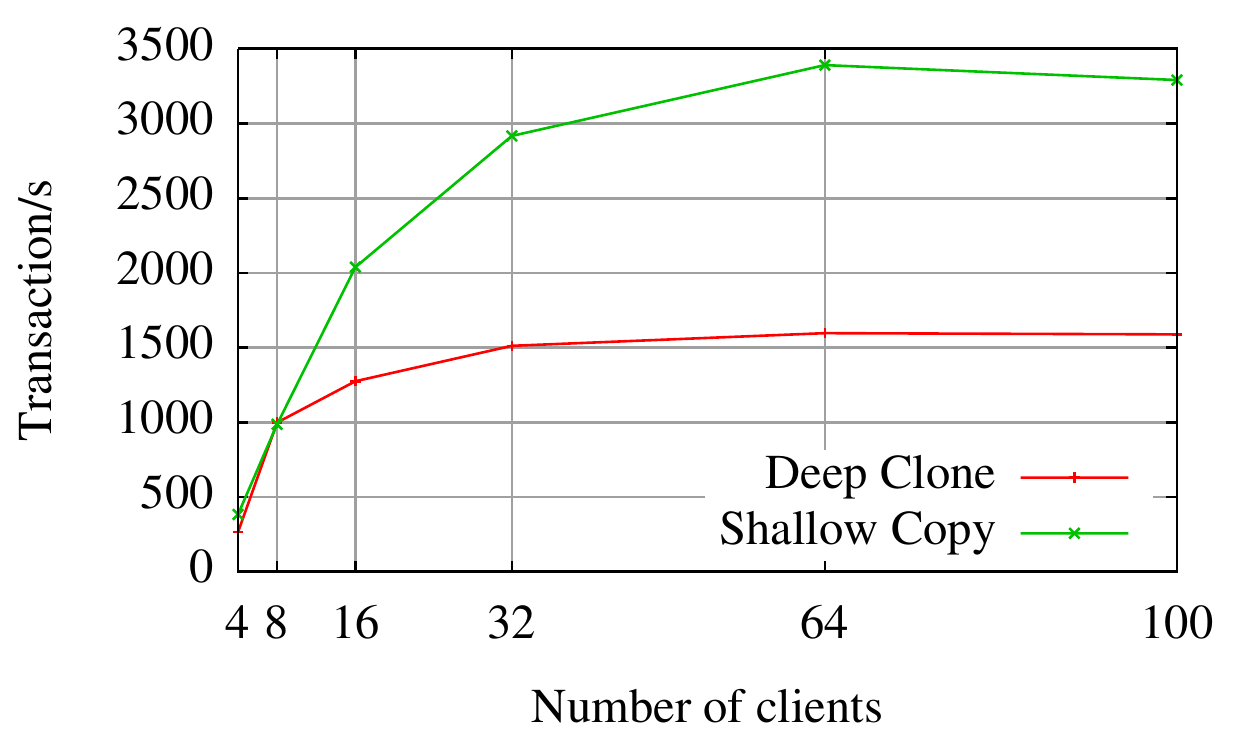}
    \caption{Cloning of Entities, comparing deep and shallow clones}
    \end{subfigure}
    \begin{subfigure}[b]{0.49\textwidth}
    \centering
    \includegraphics[width=\textwidth]{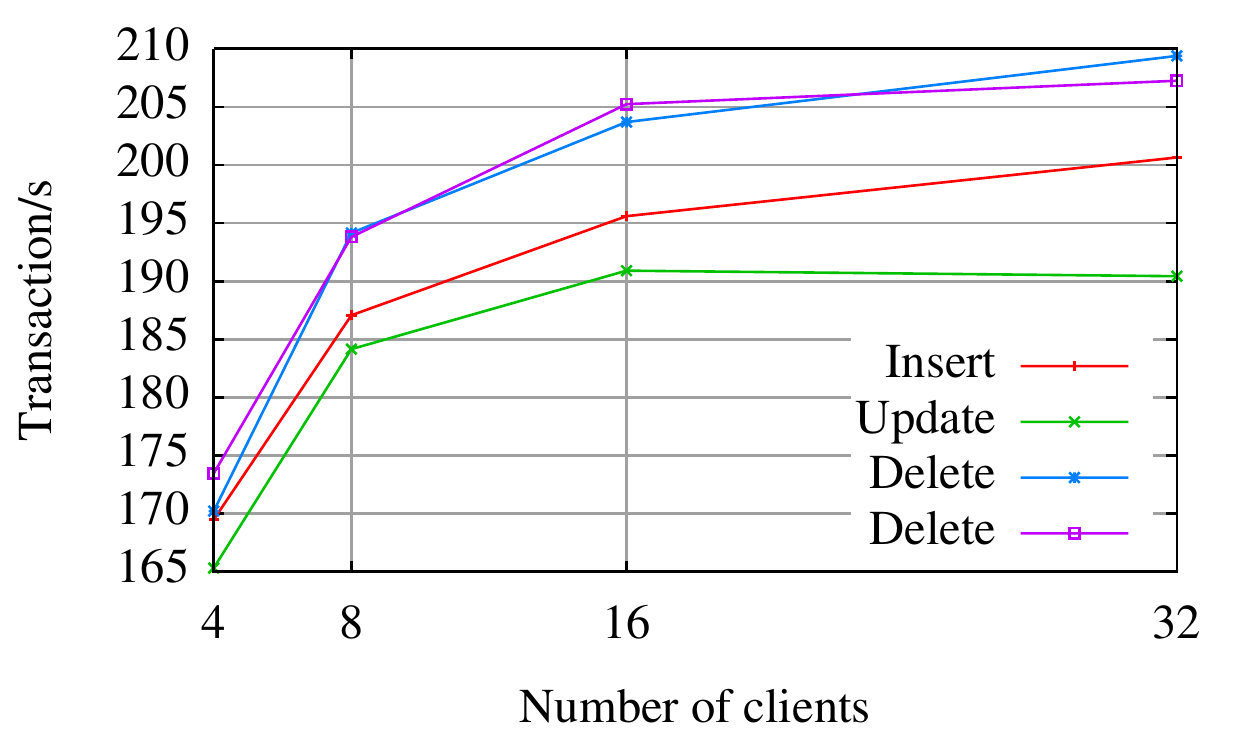}
    \caption{Throughput with RaspberryPI as bridge}
    \end{subfigure}
  \end{center}
  \caption{Throughput for cloning operations and throughput for using a RaspberryPi bridge}
  \label{fig:figure_tx_part2}
\end{figure}

Another important operation is cloning an entity. There
exist two different kinds of clone operations: a shallow copy and a deep clone.
While the shallow copy basically only inserts a single link between the old and
the new entity, the deep copy entirely copies the
current version of the set of properties and links to the new entity. Figure \ref{fig:figure_tx_part2} 
compares the results. As
expected, the shallow copy outperforms the deep copy by a factor of 2-3. 
Figure \ref{fig:figure_tx_part2} also shows the throughput we obtain using a very cheap device (a RaspberryPI) as a bridge. 

\subsubsection{Transaction Rate vs Conflict Rate}

\begin{figure}[h!tb]
  \begin{center}
    \begin{subfigure}[b]{\textwidth}
      \centering
      \includegraphics[width=.7\textwidth]{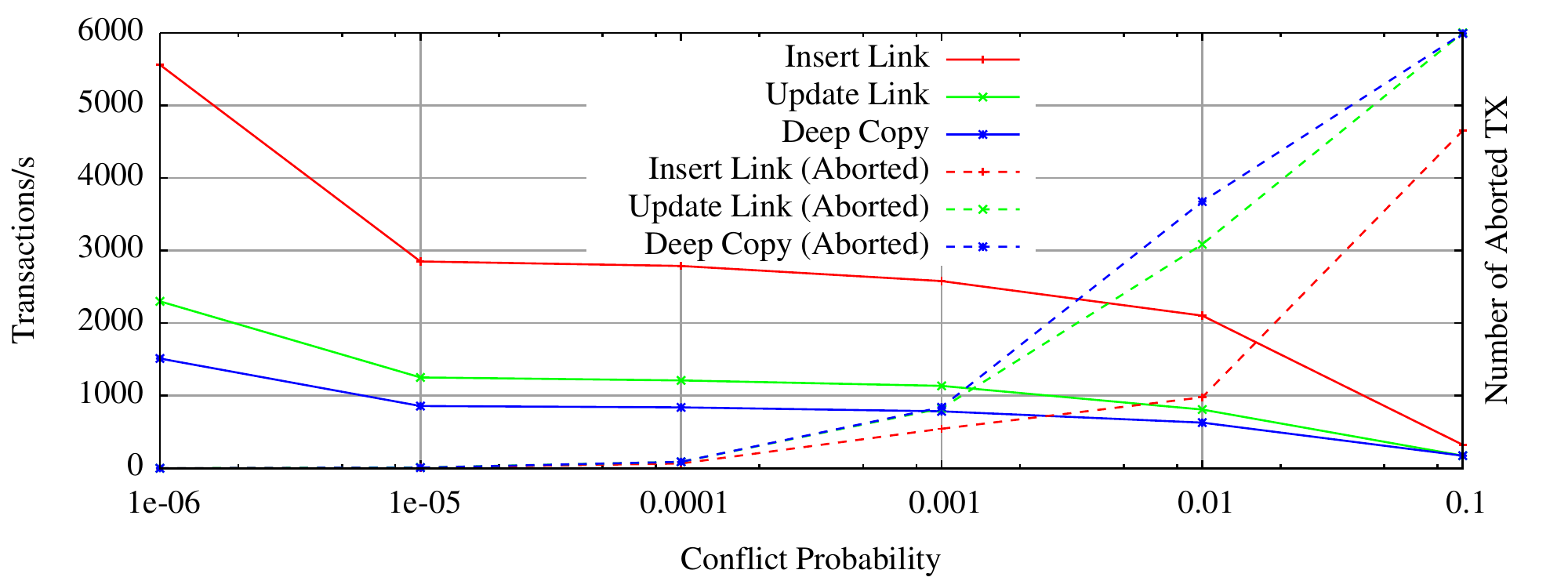}
      \caption{Throughput vs Conflict Rate}
      \label{fig:tx_conf}
    \end{subfigure}
  \end{center}

  \begin{center}
    \begin{subfigure}[b]{\textwidth}
      \centering
      \includegraphics[width=.7\textwidth]{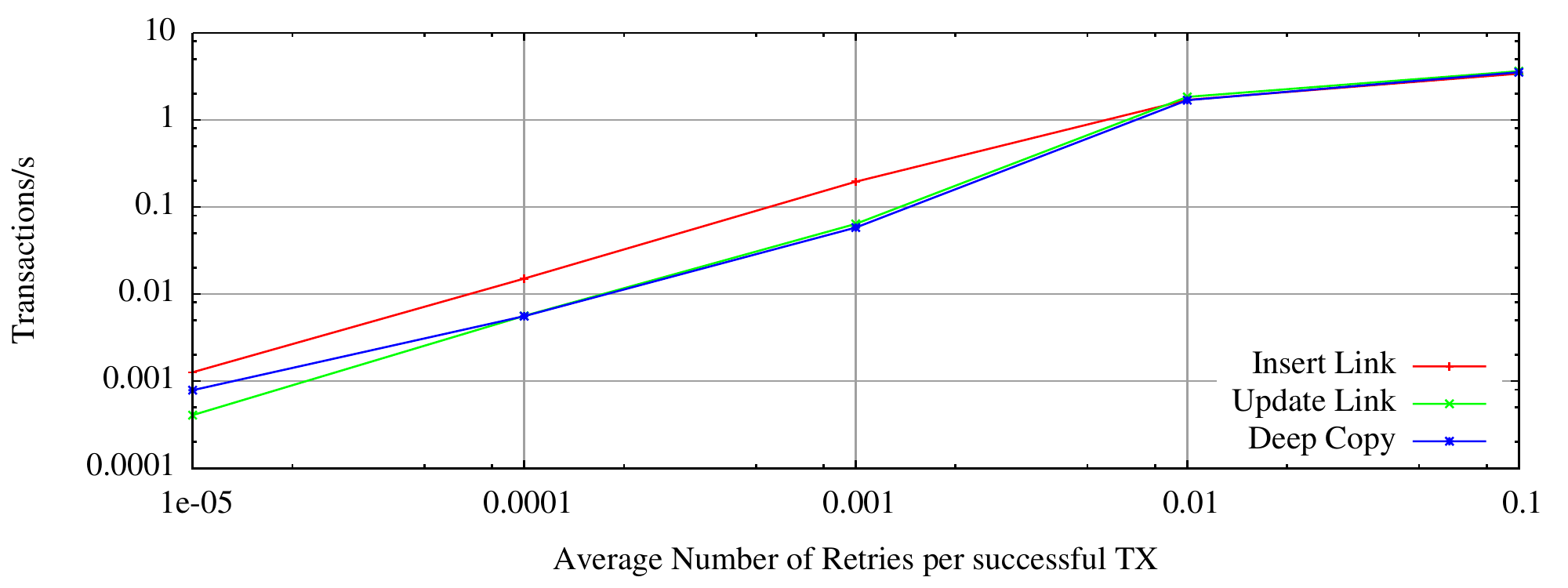}
      \caption{Number of Retries vs Conflict Rate}
      \label{fig:tx_retries}
    \end{subfigure}
  \end{center}

  \caption{Impact of conflicts on transaction throughput.
  Figure~\ref{fig:tx_conf} shows the with increasing probability of a locking
  conflict and Figure~\ref{fig:tx_retries} shows the number of retries per
  successful transaction.}
  \label{fig:figure_tx_part3}
\end{figure}

In a next series of experiments, we wish to observe the impact of our locking
strategy on the overall transaction throughput. Therefore, we use 32 parallel
clients, each executing 10k transactions sequentially. To simulate
conflicts between two clients, every client uses a specified list of
input entities to perform either an insert of a new link, an update or a delete
of a link. From the original number of 1M entities, we decrease the number of
entities to choose from equally for each client. Therefore, the probability of a
conflict increases. Figure~\ref{fig:figure_tx_part3} shows the results of this
experiment. The top Figure~\ref{fig:tx_conf} shows the transaction throughput
when increasing number of conflicts. As expected, the throughput decreases, as
it essentially serializes when only a single entity is used.

Figure~\ref{fig:tx_retries} analyzes the average number of retries a successful
transactions needs before it can be executed. The number of retries defines
the latency of a successful transaction. It
allows to not only compensate for transaction bottlenecks but for dropped
connections as well.

\section{Conclusions}
\label{sec:Conclusion}

In this work, we have presented and evaluated ERS, a new entity registry system
enabling the sharing and multipartite editing of entity data in poorly-connected contexts.
ERS is available as an open-source package and is currently being integrated in several environments,
including the Sugar desktop environment\footnote{\url{http://www.sugarlabs.org/}}. Our hope is
that ERS contributes to bridging the gap between highly-connected settings where LOD sharing and data-intensive
applications abound, and the rest of the world (representing several billion persons), where data connectivity cannot be taken as granted. 

\section*{Acknowledgment}
This work was supported by the Verisign Internet Infrastructure Grant program 
2012.

\bibliography{bib}{}

\begin{thebibliography}{1}

\bibitem{icend2011}
Christophe Gu{\'e}ret and Stefan Schlobach.
\newblock {SemanticXO}: Connecting the {XO} with the world's largest
  information network.
\newblock In JimJames Yonazi, Eliamani Sedoyeka, Ezendu Ariwa, and Eyas
  El-Qawasmeh, editors, {\em e-Technologies and Networks for Development},
  volume 171 of {\em Communications in Computer and Information Science}, pages
  261--275. Springer Berlin Heidelberg, 2011.

\bibitem{Gueret2011}
Christophe Gu\'{e}ret, Stefan Schlobach, Victor~De Boer, Anna Bon, and Hans
  Akkermans.
\newblock {Is data sharing the privilege of a few ? Bringing Linked Data to
  those without the Web}.
\newblock In {\em Proceedings of ISWC2011 - "Outrageous ideas" track, Best
  paper award}, pages 1--4. Best paper award, 2011.

\bibitem{ladwig2011cldmonks}
G{\"u}nter Ladwig and Andreas Harth.
\newblock {CumulusRDF: Linked Data Management on Nested Key-Value Stores}.
\newblock In {\em Proceedings of the 7th International Workshop on Scalable
  Semantic Web Knowledge Base Systems (SSWS2011) at the 10th International
  Semantic Web Conference (ISWC2011)}. -, Oktober 2011.

\bibitem{DBLP:conf/www/MaaliCP11}
Fadi Maali, Richard Cyganiak, and Vassilios Peristeras.
\newblock Re-using cool uris: Entity reconciliation against lod hubs.
\newblock In Christian Bizer, Tom Heath, Tim Berners-Lee, and Michael
  Hausenblas, editors, {\em LDOW}, volume 813 of {\em CEUR Workshop
  Proceedings}. CEUR-WS.org, 2011.

\end{thebibliography}
\bibliographystyle{plain}

\end{document}